# Nature of the metallic and in-gap states in Ni-doped SrTiO$_3$

Fatima Alarab[1]*, Karol Hricovini[2], Berengar Leikert[3], Christine Richter[2], Thorsten Schmitt[1], Michael Sing[3], Ralph Claessen[3], Ján Minár[4]* and Vladimir N. Strocov[1]*

[1]Swiss Light Source, Paul Scherrer Institute, 5232 Villigen-PSI, Switzerland
[2]LPMS, CY Cergy Paris University, Neuville-sur-Oise, France
[3]Physikalisches Institut and Würzburg-Dresden Cluster of Excellence ct.qmat,
Julius-Maximilians-Universität, 97074 Würzburg, Germany
[4]New Technologies Research Centre, University of West Bohemia, 301 00 Plzeň, Czech Republic
* Corresponding authors (fatima.alarab@psi.ch, jminar@ntc.zcu.cz, vladimir.strocov@psi.ch)

Epitaxial thin films of SrTiO$_3$(100) doped with 6% and 12% Ni are studied with resonant angle-resolved photoelectron spectroscopy (ARPES) at the Ti and Ni $L_{2,3}$-edges. We find that the Ni doping shifts the valence band (VB) of pristine SrTiO$_3$ towards the Fermi level (p-doping) and reduces its band gap. This is accompanied by an upward energy shift of the Ti $t_{2g}$-derived mobile electron system (MES). Thereby, the in-plane $d_{xy}$-derived bands reduce the embedded electron density, as evidenced by progressive reduction of their Fermi momentum with the Ni concentration, and the out-of-plane $d_{xz/yz}$-derived bands depopulate, making the MES purely two-dimensional. Furthermore, the Ti and Ni $L_{2,3}$-edge resonant photoemission is used to identify the Ni 3$d$ impurity state in the vicinity of the valence-band maximum, and decipher the full spectrum of the $V_O$-induced in-gap states originating from the Ni atoms, Ti atoms, and from their hybridized orbitals. Our experimental information about the dependence of the valence bands, MES and in-gap states in Ni-doped SrTiO$_3$ may help development of this material towards its device applications associated with the reduced optical band gap.

# 1 Introduction

Transition metal oxides exhibit a diversity of exotic physical properties including superconductivity, ferromagnetism, ferroelectricity, etc. [1] which can potentially be harnessed for emerging applications in electronic and quantum devices. The prototype perovskite complex oxide $SrTiO_3$ (STO) has in recent decades been a subject of intensive research both experimentally and theoretically. Quite a few fundamental quantum phenomena such as the 2-dimensional electron conductivity [2], magnetism, correlation effects and superconductivity [3-7] meet in this material, promising novel device functionalities. A novel promise of STO is its use for photocatalysis [8].

The primary limitation of STO for photovoltaic applications is its large band gap (>3 eV), which restricts light absorption to the ultraviolet range. Doping of STO with Ni (STO:Ni), where Ni substitutes Ti atoms, induces a slight reduction in the band gap, thus extending the optical absorption edge into the visible light [9,10]. Furthermore, the Ni doping introduces a spectrum of in-gap states which radically change the photoabsorption and photoluminescence properties of STO to polychromic [9]. Similar physical phenomena are known, for example, for STO doped with oxygen vacancies ($V_O$s) whereby its photoluminescence spectrum can be varied from red [11] to blue [12]. Further prospects for STO:Ni can be connected with the Rashba splitting at the surfaces and interfaces of STO-based materials [13-18]. This phenomenon, particularly pronounced at the LAO/STO interfaces, may find practical application for the spin-charge interconversion through the Edelstein and inverse Edelstein effects [19-21]. The Ni-doping can affect this interconversion through breaking of the Kramers degeneracy via the magnetic field and through modulation of the spin-orbit coupling field interplaying with the Rashba splitting.

The nature of the electronic states in STO:Ni and their evolution upon increase of the Ni doping still remains controversial. For instance, our Density-Functional theory (DFT) calculations of the partial density of states (PDOS) reveal an energy overlap between the Ti-derived and Ni-derived states in the STO band gap, which has been confirmed by resonant photoemission (ResPE) data at the Ti and Ni $L_{2,3}$-edges [10]. At the same time, the formation of the MES at the surface of STO:Ni and the effects of Ni doping on its fundamental electronic properties such as the band order and band filling have not yet been explored, to the best of our knowledge. There are many other aspects of the Ni-doping influence on the fundamental electronic structure of STO, including the in-gap states, which still remain open.

Here, we aim to understand the influence of Ni doping on the electronic structure of STO:Ni thin films ranging from the VB to the continuum of the in-gap states and to the MES in the vicinity of the Fermi level ($E_F$). By employing Ti and Ni $L_{2,3}$-edge resonant ARPES, we attempt to explore the elemental character and orbital hybridization through the whole electronic structure. Our investigation includes STO(100) single crystal as the pristine reference material.

# 2 Experimental details

We employed the pulsed laser deposition (PLD) technique to achieve epitaxial growth of STO:$Ni_x$ films with varying Ni concentrations ($x$=0.06 and 0.12). A principal challenge in fabricating STO:Ni films lies in the successful substitution of Ni cations into the STO lattice sites. The calculated formation energy for STO:Ni reveals a preference for Ni atoms to substitute Ti cations (7.8 eV for Ni at Ti sites) over Sr cations (9 eV for Ni at Sr sites) [22]. However, this value remains quite high due to the very limited solubility of Ni within STO, imparting experimental complexities and favoring the formation of metal clusters within the host lattice [23]. To mitigate Ni clustering and promote oxidation towards the $Ni^{x+}$ state (2<x<4), we employed mixed powder $[SrTiO3]_{1-x}/[NiO]_x$ target with $x$=6 at.% and 12 at.%. These films were deposited onto $TiO_2$-terminated (100) surfaces of niobium-doped (5 at.%) STO substrates. Our optimization process identified substrate temperatures of 700°C, laser energy of 1.8 J/cm$^2$, and an oxygen background gas pressure of 1·10⁻¹ mbar. Deposition was ceased after growing 11 unit cells on each substrate. The structural characterization data of the films have been detailed in [10], confirming the successful substitution of Ni ions into the Ti sites of STO.

Considering the ex-situ transfer of all samples to the experimental endstation, a surface cleaning protocol was inquired to mitigate surface contamination. This involved ozone etching of the samples, followed by thermal annealing in a vacuum of 2·10⁻⁹ mbar at 300°C for 30 minutes. The same treatment was applied to both the STO:Ni films and the bare STO substrate, serving as our reference sample.

X-ray absorption (XAS), ResPE, and ARPES measurements were conducted at the soft X-ray ARPES endstation of the ADRESS beamline at the Swiss Light Source [24,25]. Circularly polarized incident light was employed. The analyzer slit was oriented in the plane formed by the incident light and the surface normal (for details of the experimental geometry see [24]). The spot size on the sample was around ~30x75 μm. Ultrahigh vacuum of ~1·10⁻¹⁰ mbar and the sample temperature of 12 K were maintained during the experiment. As the X-ray irradiation can generate $V_O$s in STO-based samples, the measurements were performed at saturation of the spectra after ~1 hour of irradiation. For all experimental ARPES images presented, the coherent spectral fraction, reduced by the Debye-Waller factor [26], was enhanced by subtracting the angle-integrated spectrum. The ResPE spectra were acquired at the Ti and Ni $L_{2,3}$-edges ($2p \rightarrow 3d$ resonances) to selectively enhance the signal originating from the Ti- and Ni-derived electron states, respectively. The normal-emission angle (Γ-point in the $k_x$ direction along the analyzer slit) was set from the $d_{xy}$ dispersions measured at a Ti-resonance excitation energy of 457.4 eV.

# 3 Results and Discussion

## 3.1 Ti $L_{2,3}$-edge resonant photoemission

Bulk STO is a band insulator with a large band gap of 3.2 eV. Under X-ray irradiation, however, it develops surface conductivity where the Ti 3d $t_{2g}$ derived conduction-band (CB) states become populated to form the MES. In presence of the surface breaking the degeneracy of the bulk $t_{2g}$ states, they split into the ones derived from the in-plane $d_{xy}$ orbital and the ones derived from the out-of-plane $d_{xz/yz}$ orbitals (for brevity, we will neglect the cubic to tetragonal phase transition in STO at 105 K, because the corresponding atomic displacements are relatively small [27], and keep using the cubic notation for symmetry). It is generally accepted that the formation of this MES is attributed to formation of $V_O$s. In this case one of the two electrons released by the Ti atom to join the MES, and another stays localized at the Ti atom to form an in-gap state. The latter is derived from the Ti 3d $e_g$ states shifted down in energy due to strong electron correlations (for a detailed picture see [28-30] and the references therein). A qualitative difference of STO:Ni from its pristine counterpart is that it already contains Ni derived in-gap states [10] and the $V_O$s generated by X-ray irradiation only modify them and induce additional Ti 3d $e_g$ derived ones.

To capture the nature of the MES and in-gap states in STO and STO:Ni, we have performed ResPE measurements through the Ti and Ni $L_{2,3}$-edge resonances. Fig. 1(a,b) displays our results for the former as ARPES intensity maps $I(E_B,hv)$ for both STO crystal and STO:Ni$_{0.06}$ films, plotted as a function of binding energy ($E_B$) and excitation energy ($hv$). In this map, $I(E_B,hv)$ was integrated within $k_x$ = ±0.1 Å⁻¹ in order to enhance the signal from the MES states centered at Γ compared to integration over the whole angular acceptance of the analyzer.

For pristine STO, Fig. 1(a), the ResPE map reveals the resonating Ti 3d weight distributed over (1) the VB, composed of the O 2p states hybridized with Ti 3d, (2) the in-gap states, with a well-defined Ti 3d impurity state at $E_B$ ~ -1.15 eV derived from the Ti 3d $e_g$ states brought down in energy because of strong correlations [28-30], and (3) the MES in the vicinity of $E_F$, derived from the weakly correlated Ti 3d $t_{2g}$ states [28-30]. The resonance of the $t_{2g}$-derived MES appears at somewhat higher $hv$ compared to the $t_{2g}$ peak in XAS, which can be explained by remnant **k**-conservation in the ResPE process [28].

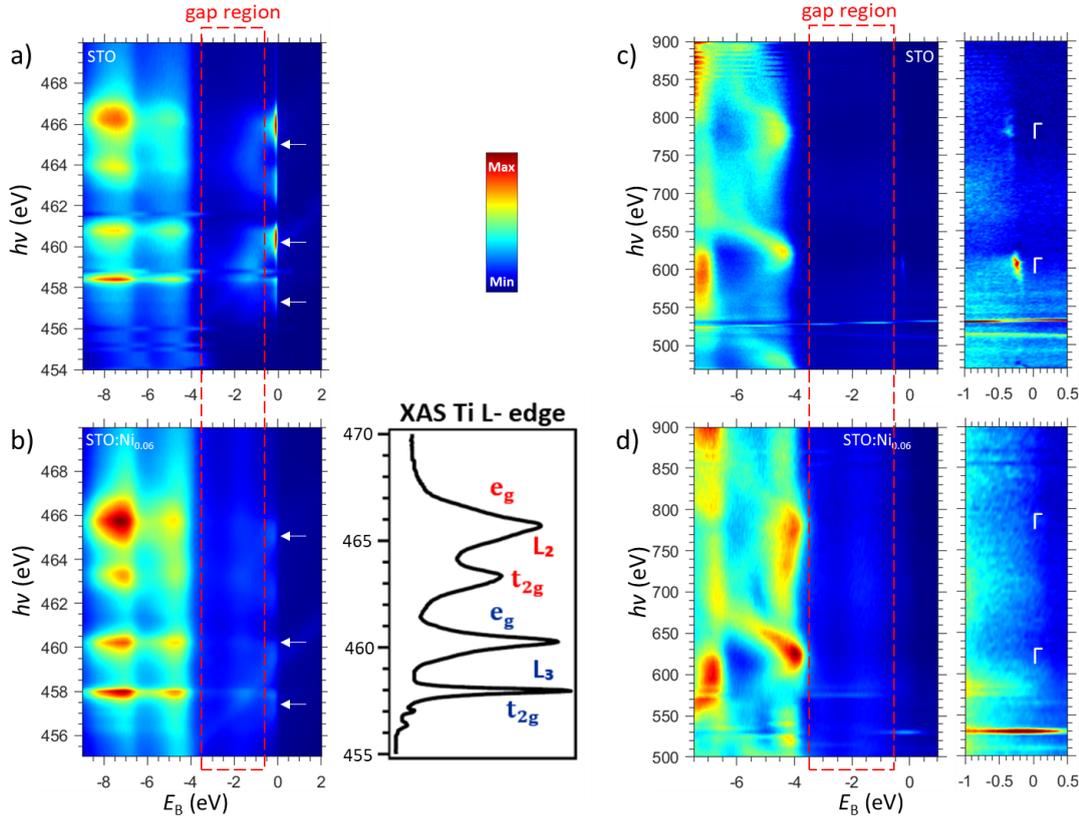

**Fig 1: $hv$-dependent ARPES of STO and STO:Ni**. (a,b) ResPE intensity map in the VB of (a) pristine STO, and (b) STO:Ni$_{0.06}$ films for $hv$ varying through the Ti $L_{2,3}$-edge, with the corresponding XAS spectrum. The white arrows indicate the $hv$ values at which the ARPES maps in Fig. 2 are taken; (c,d) $hv$ scan in an extended range above the Ti resonance, showing the out-of-plane band dispersions of (c) STO and (d) STO:Ni$_{0.06}$ films (the narrow and weak Ni $L_{2,3}$-edge resonance is not well seen). Also displayed are their zoom-in around $E_F$, showing the MES signal blowing up at $k_z$ in the Γ-points.

Turning to STO:Ni$_{0.06}$ in Fig. 1(b), the Ti $L_{2,3}$-edge XAS spectrum is hardly distinguishable from that of pristine STO [28-30]. With its main peaks labeled, the spectrum exhibits the same features in the STO case [32,33], the spin-orbit splitting of the $L_3$- and $L_2$-edges and the crystal-field splitting into $t_{2g}$ and $e_g$ orbitals. In the ResPE map, however, we observe that the Ti 3$d$ impurity state is shifted from $E_B$ = -1.15 eV to -1.4 eV. Furthermore, the whole in-gap spectral intensity and the MES signal significantly reduce in comparison with the STO case. As we will see below, the latter is consistent with the depopulation of the $d_{xz/yz}$ states penetrating deeper into the STO bulk compared to the $d_{xy}$ ones located closer to the surface.

## 3.2 Three-dimensional electronic structure

The map in Fig. 1 (c,d) covers an extended $hv$ region above the Ti $L_{2,3}$-edge resonance. Again, $I(E_B,hv)$ in this map was integrated within $k_x$ = ±0.1 Å$^{-1}$. The dispersing ARPES peaks reflect the VB dispersions as a function of the out-of-plane momentum $k_z$. Turning to the MES signal in the STO map, Fig. 1(c), we note that in the shown $hv$ region above 500 eV the photoelectron inelastic mean free path $\lambda_{PE}$ is relatively large. Therefore, the MES signal here comes predominantly from the out-of-plane $d_{xz/yz}$ states, because their larger extension into the bulk compared to the in-plane $d_{xy}$ derived ones provides larger overlap with the final state extending over $\lambda_{PE}$. The signal from the MES appears at $hv$ = 605 and 780 eV, where $k_z$ is around the Γ-points [31]. This observation is consistent with the three-dimensional nature of the $V_O$-induced $d_{xz/yz}$ states. We also notice that the Ni doping shifts the VB by ~0.5 eV towards $E_F$ (p-doping) and reduces the band gap by ~0.4 eV. This shift is consistent with another effect of the Ni doping discussed below, the depopulation of the out-of-plane $d_{xz/yz}$-derived states and thus vanishing of the off-resonance MES signal.

## 3.3 Ni-concentration dependence of the mobile electron system

Ti $L_{2,3}$-edge resonant ARPES images of the VB and CB states for the pristine STO and STO:Ni samples, measured at few representative excitation energies marked in Fig. 1(a,b), are displayed in Fig. 2. We will now follow their evolution with the Ni concentration.

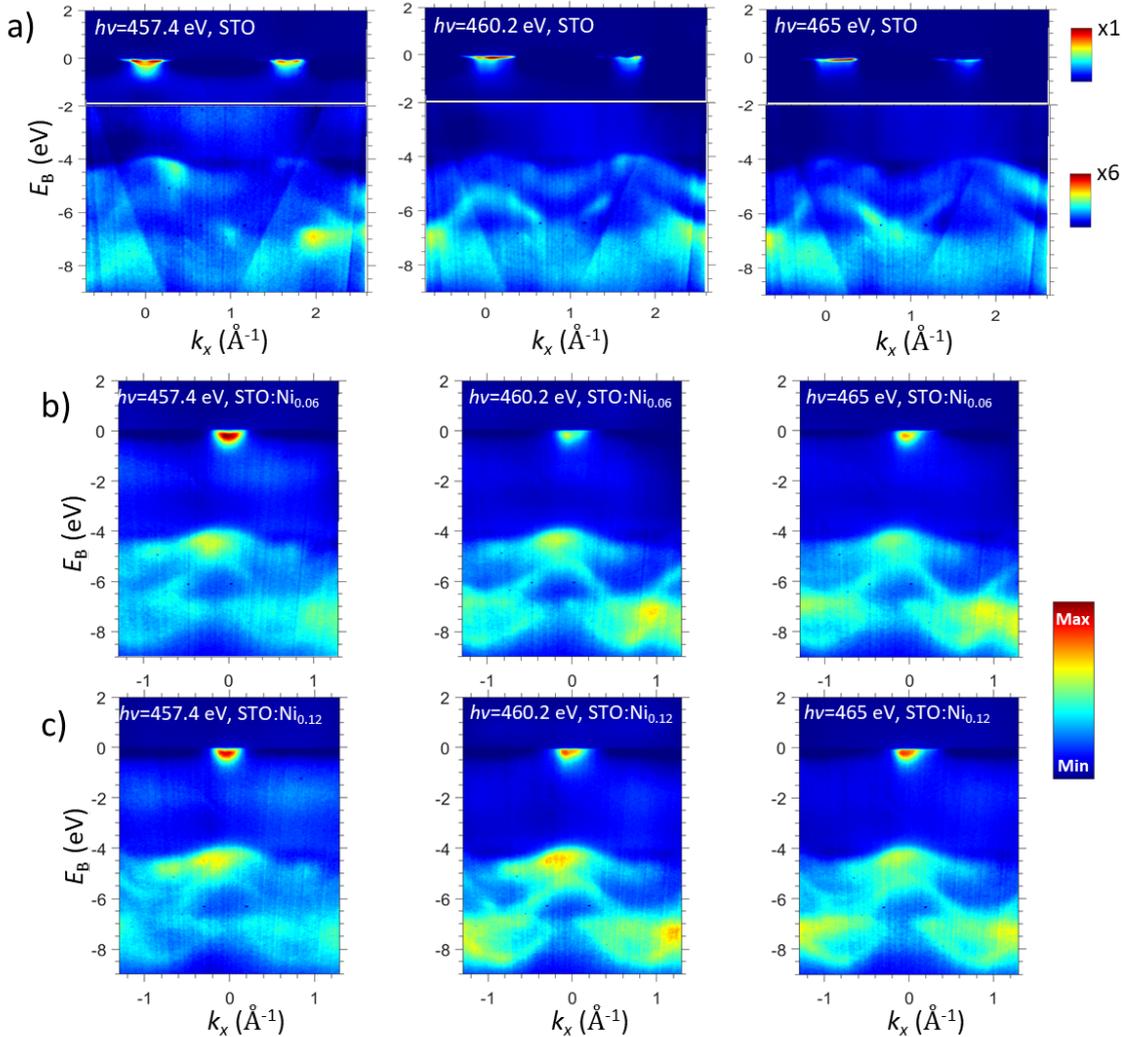

**Fig 2. Resonant ARPES** at $hv$ selected at the Ti $L_3$ ($t_{2g}$), $L_3$ ($e_g$) and $L_2$ ($e_g$) XAS peaks marked with arrows in Fig. 1(a,b): (a) STO crystal (note different intensity scale within the VB and CB regions; the inclined lines are electron optics artifacts enhanced by the background subtraction), and (b) STO:Ni$_{0.06}$ and STO:Ni$_{0.12}$ films. The MES to VB intensity ratio reduces under Ni doping because of the depopulation of the $d_{xz/yz}$ states.

For the STO sample in Fig. 2(a) and the STO:Ni$_{0.06}$ and STO:Ni$_{0.12}$ ones in (b), all maps show dispersive bulk bands located within an $E_B$ range of -4 eV to -9 eV, arising from the O 2$p$ states. The $d_{xy}$-derived and $d_{xz/yz}$-derived bands, constituting the MES in the vicinity of $E_F$, are located near the Γ-point. To effectively differentiate the light $d_{xy}$ band from the heavy $d_{yz}$ ones (see Ref. [28]) we used three $hv$ values going through the $t_{2g}$ and $e_g$ states of the $L_3$ and $L_2$ edges. The first observation is that the MES to VB intensity ratio for the STO:Ni samples is nearly an order of magnitude smaller compared to the STO one (note the different intensity scales in the MES and VB energy regions in (a)). As we will see below, the Ni doping depopulates the $d_{xz/yz}$ states, giving maximal contribution to the MES signal. The concomitant reduction of the MES intensity apparently scales the VB intensity in our ARPES images, where the color scale is normalized to the maximal intensity.

Fig. 3(a-c) zooms in the MES region of our samples, respectively. One can clearly see the light $d_{xy}$ bands and the heavy $d_{yz}$ ones (the light $d_{xz}$ bands, degenerate with the $d_{yz}$ ones in the Γ-point, are hardly visible on top of the $d_{yz}$ intensity). The first observation is that whereas the $d_{xy}$ bands are visible for all samples at all excitation energies, the heavy $d_{yz}$ bands fade away with Ni doping and can be seen only at $hv$ = 465 eV. This behavior indicates gradual shifting of the MES states to higher energies by ~0.1 eV which leads, in the first place, to depopulation of the $d_{yz}$ bands. This energy shift is confirmed by analysis of the $k_F$ values of the $d_{xy}$ and $d_{yz}$ bands in relation to Ni content. We have extracted these values from the momentum-distribution-curve (MDC) cuts of the spectral intensity $W(k)$ integrated within a 50 meV range around $E_F$ for each case, Fig. 3(c). The extremes of their gradient $dW/dk$ identify the $k_F$ values [31,34,35]. First, we see that the $k_F$ signatures of the $d_{yz}$ bands $k_F(d_{yz})$, clear for pristine STO, disappear upon Ni doping. We may only notice a remnant weight of the $d_{yz}$ band surviving at $hv$ = 465 eV (where the $d_{yz}/d_{xy}$ intensity ratio is maximal [28]) with $k_F$ of this band being close to that of pristine STO; this fact may indicate fluctuations of the Ni concentration in our STO:Ni samples where, in the spirit of the phase-separation picture, a small volumetric fraction staying nearly pristine. Focusing on the $d_{yz}$ bands, we find their $k_F$ values $k_F(d_{xy})$ systematically decrease from 0.20 Å$^{-1}$ in pristine STO to 0.15 Å$^{-1}$ upon 12% Ni doping. This behavior of the $d_{xy}$ bands is fully consistent with the shift of the MES to higher energies and depopulation of the $d_{xz/yz}$ ones. We note that the depopulation fully develops already the smallest Ni concentration of 6%.

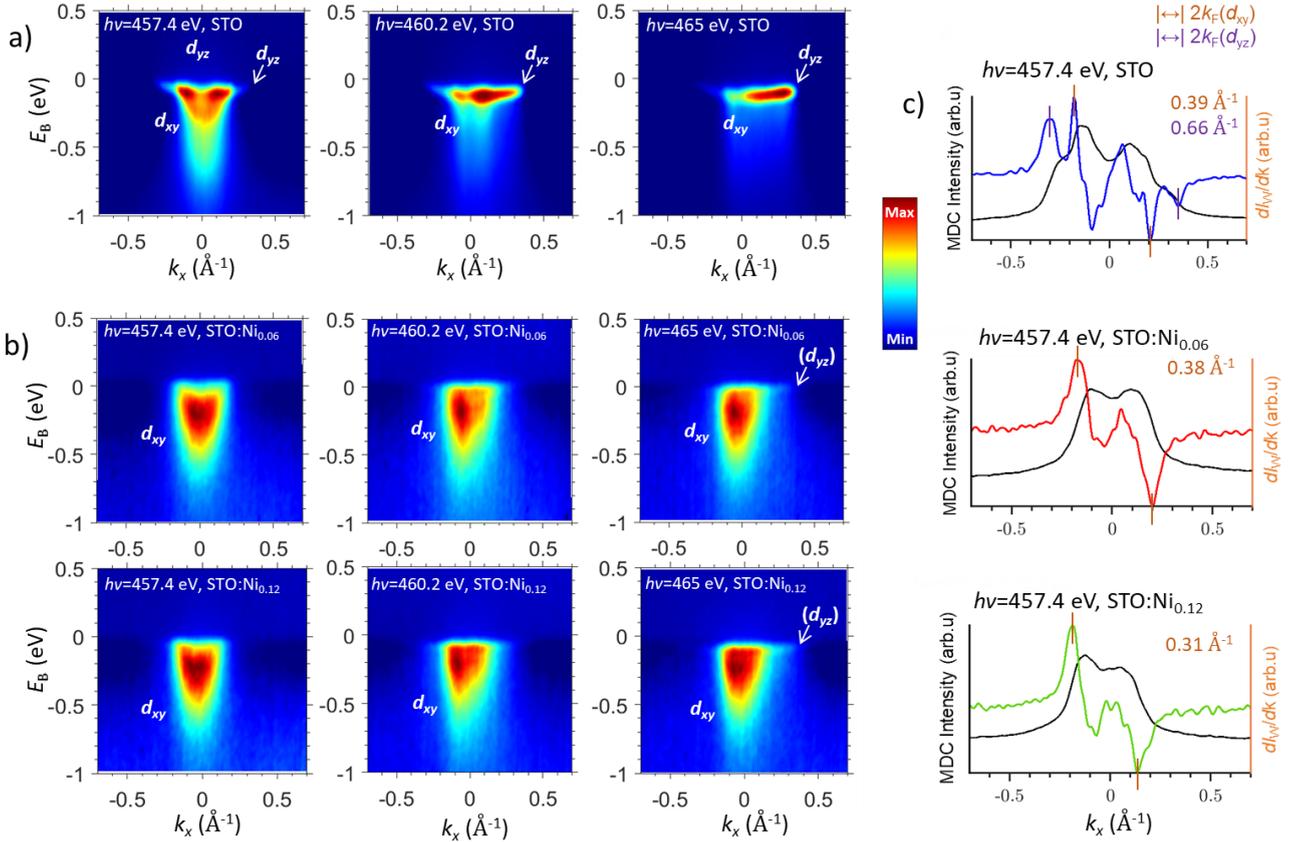

**Fig. 3. Effect of Ni-doping in resonant ARPES of the MES**: (a) STO crystal and (b) STO:Ni$_{0.06}$ and STO:Ni$_{0.12}$ films (zoom-in of the data in Fig. 2). (c) MDC cuts at $hv$ = 457.4 eV (integrated within $E_F$ ± 50 meV) and the corresponding gradient $dW/dk$ (color lines) of the bandwidth-integrated intensity $W(k)$ (black). The maxima/minima of $dW/dk$, marked by the red and purple vertical lines, identify the $k_F$ values for the $d_{xy}$ and $d_{yz}$ bands, whose numerical values are displayed in the corresponding colors.

Another interesting observation from Fig. 3(a-c) is that the MES spectral intensity is notably asymmetric relative to the Γ-point. Such an asymmetry is principle not prohibited because the angle between the surface normal and the X-ray polarization vector light is different for $k_x$ on the opposite sides of Γ; another contribution to the asymmetry in the photon momentum which is ~0.24 Å$^{-1}$ at the Ti $L_{2,3}$-edge [31]. This effect should trace back to the different behavior of the $d_{xy}$ and $d_{xz/yz}$ bands across the resonance, vividly discussed in Ref. [28], combined with their gradual depopulation with increase of the Ni doping.

## 3.4 Ni $L_{2,3}$-edge resonant photoemission

In our previous work [10], the origin of the main VB features of the Ni-doped STO films were investigated both experimentally and theoretically by means of (angle-integrated) ResPE and DFT calculations, respectively. In the present work, we deepen this analysis based on a detailed ResPE map $I(E_B,hv)$ presented in Fig. 4 (a,c). Here, the VB spectra of STO:Ni$_{0.06}$ and STO:Ni$_{0.12}$ are recorded across the Ni $L_3$ absorption edge with $hv$ varying from 850 eV to 856 eV in increments of 0.5 eV. To accentuate the Ni resonant spectral weight, (b,d) presents a map of the differential intensity $\Delta_{res}(E_B,hv) = I(E_B,hv) - I(E_B,hv_0)$ obtained by subtracting, from each $I(E_B,hv)$ spectrum, the pre-resonance one $I(E_B,hv_0)$ measured at $hv_0 =$ 850 eV.

Beginning from the XAS spectrum for the STO:Ni$_{0.06}$ case in Fig. 4 (a), we note that the main Ni $L_3$ absorption edge, situated at 853.3 eV, is followed by a weaker peak around 855 eV. The latter is attributed to multiplet splitting of the $2p^53d^9$ state, connected with the strongly-correlated nature of Ni-ions in the crystal field [10]. In the corresponding ResPE map across the Ni $L_3$ edge ($hv$ from 852.5 to 853.5 eV), we observe, first of all, an intensity boost of the whole VB, indicating the hybridization of the O 2$p$ states with the Ni 3$d$ ones. Furthermore, we observe a notable resonance of the whole in-gap spectral range down to the VB maximum (VBM). This resonant behavior is accentuated in the $\Delta_{res}$ map (b). Here, we can discern two distinct resonant peaks, where the first manifests the main Ni 3$d$ impurity state at $E_B$ of -3.5 eV (~0.5 eV above the VBM) [36] and the second a Ni-derived in-gap state at -1.9 eV. While with increase of $hv$ all these resonant structures stay at constant $E_B$, which is characteristic of the coherent resonant photoemission process, we note a feature near the VB bottom which appears at $hv$ above the VB resonance and follows a constant-kinetic-energy line. This feature is associated with an incoherent process of resonant Auger decay [3].

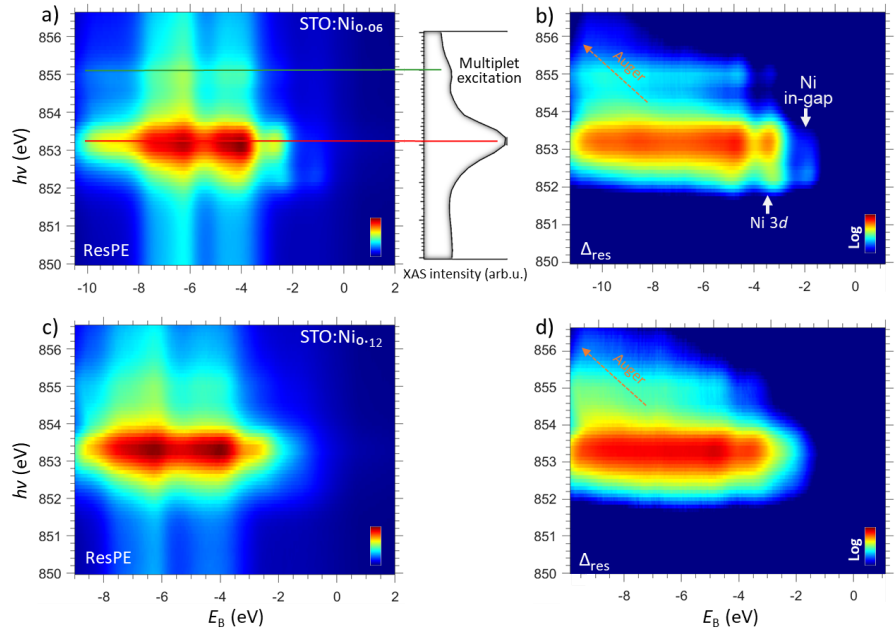

**Fig. 4: Ni $L_3$-edge ResPE intensity data map** for (a) STO:Ni$_{0.06}$ and (c) STO:Ni$_{0.12}$ films, with the $L_3$-edge XAS spectrum for STO:Ni$_{0.06}$ shown in (a); (b,d) Differential ResPE maps (logarithmic scale), where the pre-resonant spectrum at $hv_0$=850 eV is subtracted, for STO:Ni$_{0.06}$ (b) and STO:Ni$_{0.12}$ (d). These maps show the resonant contribution of the Ni states $\Delta_{res}$ to the VB spectral weight.

The experimental ResPE map for the STO:Ni$_{0.12}$ films is presented in Fig. 4 (c,d). As expected, the increase of the Ni content produces more pronounced and broader resonant peaks compared to the

STO:Ni$_{0.06}$ case. The in-gap state at -1.9 eV is smeared out with the Ni concentration increase. Otherwise, the overall ResPE intensity pattern closely resembles that of STO:Ni$_{0.06}$.

## 3.5 Ti- and Ni-derived in-gap states

In the last section, we will focus on possible origins of in-gap states observed in our STO:Ni samples. Fig. 5(a) presents the (angle-integrated) ResPE spectra in the in-gap region measured at the Ti($e_g$) and Ni $L_3$ resonances at $h\nu$ = 460 and 853 eV, respectively. In STO, the in-gap states resonate at the Ti edge at $E_B$ = -1.15 eV, similarly to the previous reports [2, 28, 35]. These in-gap states originate from Ti ions located near the V$_O$s, and are often viewed as small polarons [28]. Upon 6% Ni doping, this resonant peak broadens and shifts to $E_B$ = -1.4 eV, suggesting hybridization of these states with the Ni 3$d$ ones in STO:Ni. At the Ni resonance, we observe the Ni 3$d$ impurity state at $E_B$ ~ -3.5 eV and the in-gap state -1.9 eV. This observation consequently suggests the presence of two distinct types of in-gap states in STO:Ni samples. The first type, resonating at the Ti $L_3$ edge, corresponds to localized charge carriers from the Ti ions near the V$_O$s, which may hybridize with the Ni dopants. The second type of the in-gap states, positioned at -1.9 eV, resonate only at the Ni $L_{2,3}$-edge. Based on the similarity of its energy with the Ti derived in-gap electrons, we attribute them to the V$_O$s formed within Ni octahedra.

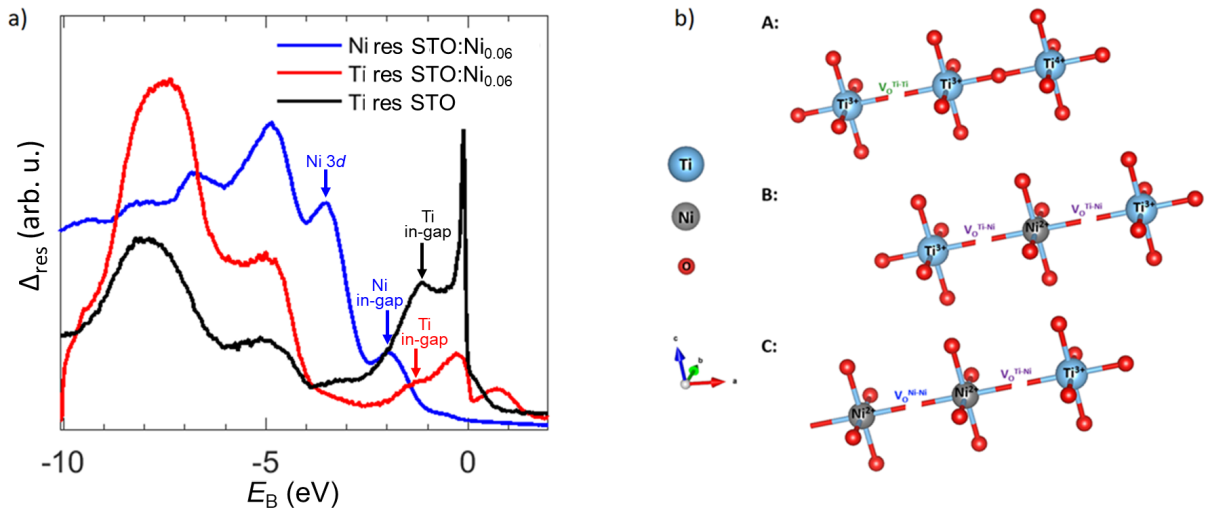

**Fig. 5: In-gap states and their orbital configurations:** (a) Resonating Fermi states and the in-gap states of (black) pristine STO crystal and (red) STO:Ni$_{0.06}$ films recorded at the Ti $L_3$($e_g$) resonance at $h\nu$ = 460 eV, and of (blue) STO:Ni$_{0.06}$ films recorded at the Ni $L_3$ resonance at $h\nu$ = 853 eV. (b) Atomic configurations of the octahedral in STO:Ni, showing different incorporation of the V$_O$s relative to the Ti and Ni atoms leading to different in-gap states.

Fig. 5(b) illustrates different variants of atomic configurations of the octahedra in STO:Ni which can incorporate V$_O$s and subsequently generate the diverse in-gap states discussed above. The scheme A reflects the scenario of V$_O$s linked to Ti$^{3+}$ species, akin to those in pristine STO. Alternatively, the scheme B shows their association with both Ti$^{3+}$ and Ni$^{2+}$ ions. Another configuration, outlined in the scheme C, shows V$_O$s surrounded solely by Ni$^{2+}$ ions, ultimately leading to entirely different in-gap states at $E_B$ = -1.9 eV. The electronic configurations and their connection with the electron paramagnetic resonance and photochromic optical absorption for this configuration were discussed in detail by Koidl et al. [9].

# 6 Conclusion

We have explored the delocalized MES and localized in-gap states in STO:Ni$_{0.06}$ and STO:Ni$_{0.12}$ films in comparison with pristine STO using resonant soft-X-ray ARPES at the Ti and Ni $L_{2,3}$-edges. Our main findings include: (1) The 3D band structure of STO:Ni and STO, where the Ni doping shifts the VB by

~0.5 eV towards $E_F$ (p-doping) and reduces the band gap by ~0.4 eV; (2) The doping shifts the CB-derived MES states to higher energies by ~0.1 eV. This shift reduces electron density in the in-plane $d_{xy}$ states, as evidenced by gradual reduction of the corresponding $k_F$ with increase of the Ni concentration, and completely depopulates the out-of-plane $d_{xz/yz}$ states, with a faint remnant weight giving an inkling of Ni-doping inhomogeneity. The prevalence of the $d_{xy}$ states renders the MES at most two-dimensional; (3) The Ni 3$d$ impurity state has been identified at ~0.5 eV above the VBM. Furthermore, the full spectrum of the $V_O$-induced in-gap states has been identified, including the localized electrons trapped in the 3$d$ orbitals of Ni atoms and of Ti ones. The latter shift by ~0.25 eV upon the Ni doping as the result of their hybridization with the neighboring Ni atoms. We conjecture that the above modifications of the MES and in-gap electronic spectrum are largely affected by deformation of the Ti octahedra under the lattice strain induced by the Ni doping. The knowledge of the electronic spectrum of the MES and in-gap states in STO:Ni as a function of Ni concentration, achieved in this work, may help tailoring properties of the STO:Ni-based materials to potential device applications.

# Acknowledgments

F.A. acknowledges the financial support from the Swiss National Science Foundation within the grant 200020B_188709. This publication was supported by the project QM4ST with reg. no. CZ.02.01.01/00/22_008/0004572, co-funded by the ERDF as part of the MŠMT. The work in Würzburg was supported by the Deutsche Forschungsgemeinschaft (DFG, German Research Foundation) under Germany's Excellence Strategy through the Würzburg-Dresden Cluster of Excellence on Complexity and Topology in Quantum Matter ct.qmat (EXC 2147, Project ID 390858490) as well as through the Collaborative Research Center SFB 1170 ToCoTronics (Project ID 258499086).

# References

[1] Ueno, K., Nakamura, S., Shimotani, H., Ohtomo, A., Kimura, N., Nojima, T., Aoki, H., Iwasa, Y. & M. Kawasaki Kawasaki, M. (2008). Electric-field-induced superconductivity in an insulator. Nature Materials **7**, 855−858.

[2] Santander-Syro, A. F., Copie, O., Kondo, T., Fortuna, F., Pailhès, S., Weht, R., Qiu, X. G., Bertran, F., Nicolaou, A., Taleb-Ibrahimi, A., Le Fèvre, P., Herranz, G., Bibes, M., Reyren, N., Apertet, Y., Lecoeur, P., Barthélémy, A. & Rozenberg, M. J. (2011). Two-dimensional electron gas with universal subbands at the surface of $SrTiO_3$. Nature **469**, 189−193.

[3] Gastiasoro, M. N., Ruhman, J. & Fernandes, R. M. (2020). Superconductivity in dilute $SrTiO_3$: A review. Annals of Physics **417**, 168107.

[4] Sing, M., Jeschke, H. O., Lechermann, F., Valentí, R. & Claessen, R. (2017). Influence of oxygen vacancies on two-dimensional electron systems at $SrTiO_3$-based interfaces and surfaces. The European Physical Journal Special Topics **226**, 2457−2475.

[5] Ahadi, K., Galletti, L., Li, Y., Salmani-Rezaie, S., Wu, W. & Stemmer, S. (2019). Enhancing superconductivity in $SrTiO_3$ films with strain. Science Advances **5**, eaaw0120

[6] Salmani-Rezaie, S., Galletti, L., Schumann, T., Russell, R., Jeong, H., Li, Y., Harter, J.W. and Stemmer, S. (2021). Superconductivity in magnetically doped $SrTiO_3$. Applied Physics Letters **118**, 202602.

[7] Scheerer, G., Boselli, M., Pulmannova, D., Rischau, C. W., Waelchli, A., Gariglio, S., Giannini, E., van der Marel, D. and Triscone, J.-M. (2020). Ferroelectricity, Superconductivity, and $SrTiO_3$ - Passions of K.A. Müller. Condensed Matter **5**, 60.

[8] Bera A., Wu K., Sheikh A., Alarousu E., Mohammed O.F., and Wu T. (2014). Perovskite Oxide SrTiO$_3$ as an efficient Electron Transporter for Hybrid Perovskite Solar Cells. The Journal of Physical Chemistry C **118**, 28494-28501.

[9] Koidl P., Blazey K. W., Berlinger W., and Müller K. A. (1976). Photochromism in Ni-doped SrTiO$_3$. Phys. Rev. B **14**, 2703.

[10] Alarab, F., Hricovini, K., Leikert, B., Nicolaï, L., Fanciulli, M., Heckmann, O., Richter, C., Prušakova, L., Jansa, Z., Šutta, P., Rault, J., Lefevre, P., Sing, M., Muntwiler, M., Claessen, R., and Minár, J. (2021). Photoemission study of pristine and Ni-doped SrTiO$_3$ thin films. Physical Review B **104**, 165129.

[11] Crespillo M.L., Graham J.T., Agullo-Lopez F., Zhang Y., Weber W.J. (2018) Isolated oxygen vacancies in strontium titanate shine red: Optical identification of Ti$^{3+}$ polarons. Applied Materials Today **12**, 131-137.

[12] Kan D., Terashima, T., Kanda, R., Masuno, A., Tanaka, K., Chu, S., Kan, H., Ishizumi, A., Kanemitsu, Y., Shimakawa, Y. & Takano, M. Blue-light emission at room temperature from Ar$^+$-irradiated SrTiO$_3$ (2005) Nature Mater. **4**, 816–819

[13] Yin, C., Seiler, P., Tang, L. M. K., Leermakers, I., Lebedev, N., Zeitler, U. & Aarts, J. (2020). Tuning Rashba spin-orbit coupling at LaAlO$_3$/SrTiO$_3$ interfaces by band filling. Physical Review B **101**, 245114.

[14] Lebedev, N., Stehno, M., Rana, A., Reith, P., Gauquelin, N., Verbeeck, J., Hilgenkamp, H., Brinkman, A. and Aarts, J. (2021). Gate-tuned anomalous Hall effect driven by Rashba splitting in intermixed LaAlO$_3$/GdTiO$_3$/SrTiO$_3$. Scientific Reports **11**, 10726.

[15] Noh, S., Choe, D., Jin, H. & Yoo, J.-W. (2022). Enhancement of the Rashba Effect in a Conducting SrTiO$_3$ Surface by MoO$_3$ Capping. ACS Applied Materials and Interfaces **14**, 50280–50287.

[16] To, D. Q., Dang, T. H., Vila, L., Attané, J. P., Bibes, M. & Jaffrès, H. (2021). Spin to charge conversion at Rashba-split SrTiO3 interfaces from resonant tunneling. Physical Review Research **3**, 043170.

[17] Kim, T., Kim, S.-I., Baek, S.-H., Hong, J. & Koo, H. C. (2015). Conductance Change Induced by the Rashba Effect in the LaAlO$_3$/SrTiO$_3$ Interface. Journal of Nanoscience and Nanotechnology **15**, 8632–8636.

[18] Fischer, M. H., Raghu, S. & Kim, E.-A. (2013). Spin-orbit coupling in LaAlO3/SrTiO3 interfaces: magnetism and orbital ordering. New Journal of Physics **15**, 023022.

[19] Trama, M., Cataudella, V., Perroni, C. A., Romeo, F. & Citro, R. (2022). Tunable Spin and Orbital Edelstein Effect at (111) LaAlO$_3$/SrTiO$_3$ Interface. Nanomaterials **12**, 2494.

[20] Telesio, F., Moroni, R., Pallecchi, I., Marré, D., Vinai, G., Panaccione, G., Torelli, P., Rusponi, S., Piamonteze, S., di Gennaro, E., Khare, A., Miletto Granozio, F., and Filippetti, A. (2018). Study of equilibrium carrier transfer in LaAlO$_3$/SrTiO$_3$ from an epitaxial La$_{1-x}$Sr$_x$MnO$_3$ ferromagnetic layer. Journal of Physics Communications **2**, 025010.

[21] Johansson, A., Göbel, B., Henk, J., Bibes, M. & Mertig, I. (2021). Spin and orbital Edelstein effects in a two-dimensional electron gas: Theory and application to SrTiO3 interfaces. Physical Review Research, **3**, 013275.

[22] Dong, X.-L., Zhang, K.-H. & Xu, M.-X. (2018). First-principles study of electronic structure and magnetic properties of SrTi$_{1-x}$M$_x$O$_3$ (M = Cr, Mn, Fe, Co, or Ni). Frontiers of Physics **13**, 137106

[23] Hennes, M., Schuler, V., Weng, X., Buchwald, J., Demaille, D., Zheng, Y. & Vidal, F. (2018). Growth of vertically aligned nanowires in metal\oxide nanocomposites: Kinetic Monte-Carlo modeling versus experiments. Nanoscale **10**, 7666–7675.

[24] Strocov, V. N., Wang, X., Shi, M., Kobayashi, M., Krempasky, J., Hess, C., Schmitt, T., and Patthey, L. (2013). Soft-X-ray ARPES facility at the ADRESS beamline of the SLS: concepts, technical realisation and scientific applications. Journal of Synchrotron Radiation **21**, 32–44.

[25] Strocov, V. N., Schmitt, T., Flechsig, U., Schmidt, T., Imhof, A., Chen, Q., Raabe, J., Betemps, R., Zimoch, D., Krempasky, J., Wang, X., Grioni, M., Piazzalunga, A., and Patthey, L. (2010). High-resolution soft X-ray beamline ADRESS at the Swiss Light Source for resonant inelastic X-ray scattering and angle-resolved photoelectron spectroscopies. Journal of Synchr. Radiation **17**, 631–643.

[26] Braun J., Minár J., Mankovsky S., Strocov V. N., Brookes N. B., Plucinski L., Schneider C. M., Fadley C. S., and Ebert H. (2013 ) Phys. Rev. B **88**, 205409

[27] Wang, F. F. Y. & Gupta, K. P. Phase transformation in the oxides. (1973) Metall. Trans. **4**, 2767

[28] Chikina, A., Lechermann, F., Husanu, M.-A., Caputo, M., Cancellieri, C., Wang, X., Schmitt, T., Radovic, M. and Strocov, V. N. (2018). Orbital Ordering of the Mobile and Localized Electrons at Oxygen-Deficient $LaAlO_3/SrTiO_3$ Interfaces. ACS Nano **12**, 7927–7935.

[29] Strocov, V. N., Chikina, A., Caputo, M., Husanu, M.-A., Bisti, F., Bracher, D., Schmitt, T., Miletto Granozio, F., Vaz, C. A. F., and Lechermann, F. (2019). Electronic phase separation at $LaAlO_3/SrTiO_3$ interfaces tunable by oxygen deficiency. Phys. Rev. Materials **3**, 106001

[30] Lechermann, F., Jeschke, H. O., Kim, A. J., Backes, S. & Valentí, R. (2016). Electron dichotomy on the $SrTiO_3$ defect surface augmented by many-body effects. Physical Review B **93**, 121103.

[31] Strocov, V. N., Lechermann, F., Chikina, A., Alarab, F., Lev, L. L., Rogalev, V. A., Schmitt, T., and Husanu, M.-A. (2022). Dimensionality of mobile electrons at x-ray-irradiated $LaAlO_3/SrTiO_3$ interfaces. Electronic Structure **4**, 015003.

[32] Wu, M., Xin, H. L., Wang, J. O., Li, X. J., Yuan, X. B., Zeng, H., Zheng, J.-C. and Wang, H.-Q. (2018). Investigation of the multiplet features of SrTiO3 in X-ray absorption spectra based on configuration interaction calculations. Journal of Synchr. Radiation **25**, 777–784.

[33] Zhao, Q. & Cheng, X.-L. (2019). Study on the Ti $K$, $L_{2,3}$ and $M$ Edges of $SrTiO_3$ and $PbTiO_3$. The Journal of Physical Chemistry A **124**, 322–327.

[34] Straub Th., Claessen R., Steiner P., Hufner S., Eyert V., Many-body definition of a Fermi surface: Application to angle-resolved photoemission. (1997) Physical Review B - Condensed Matter and Materials Physics **55**, 13473-13478.

[35] Wang, Z., Walker, S. M., Tamai, A., Wang, Y., Ristic, Z., Bruno, F. Y., de la Torre A., Riccò, S., Plumb, N. C., Shi, M., Hlawenka, P., Sánchez-Barriga, J., Varykhalov, A., Kim, T. K., Hoesch, M., King, P. D. C., Meevasana, W., Diebold, U., Mesot, J., Moritz, B., Devereaux, T. P., Radovic, M., and Baumberger, F. (2016). Tailoring the nature and strength of electron-phonon interactions in the $SrTiO_3$(001) 2D electron liquid. Nature Materials **15**, 835–839.

[36] van der Laan, G., Zaanen, J., Sawatzky, G. A., Karnatak, R. & Esteva, J.-M. (1986). Comparison of x-ray absorption with x-ray photoemission of nickel dihalides and NiO. Physical Review B **33**, 4253–4263.